\documentclass[a4paper,11pt]{article}    % For preprint
\pdfoutput=1 % if your are submitting a pdflatex (i.e. if you have  images in pdf, png or jpg format)

\usepackage{jheppub}
\usepackage{epsfig}
\usepackage[T1]{fontenc} % if needed
\def\arXiv#1{\href{http://arxiv.org/abs/#1}{arXiv:#1}}
\def\arXiv#1#2{\href{http://arxiv.org/abs/#1}{arXiv:#1}}

\def\doi#1#2{\href{http://doi.org/#1}{#2}}

\def\be{\begin{eqnarray}}
\def\ee{\end{eqnarray}}
\def\bea{\begin{eqnarray}}
\def\eea{\end{eqnarray}}
\newcommand{\nn}{\nonumber}
\newcommand\para{\paragraph{}}

\def\Dslash{\,\,{\raise.15ex\hbox{/}\mkern-12mu D}}
\def\Dbarslash{\,\,{\raise.15ex\hbox{/}\mkern-12mu {\bar D}}}
\def\delslash{\,\,{\raise.15ex\hbox{/}\mkern-9mu \partial}}
\def\delbarslash{\,\,{\raise.15ex\hbox{/}\mkern-9mu {\bar\partial}}}
\def\pslash{\,\,{\raise.15ex\hbox{/}\mkern-9mu p}}
\def\calDslash{\,\,{\raise.15ex\hbox{/}\mkern-12mu {\cal D}}}

\def\lae{\mathrel{\mathop{\smash{\lower .5 ex \hbox{$\stackrel<\sim$}}}}}
\def\lae{\mathrel{\mathop{\smash{\lower .5 ex \hbox{$\stackrel>\sim$}}}}}

\title{\boldmath Momentum relaxation of holographic Weyl semimetal from massive gravity}

\author{Junkun Zhao}
\affiliation{
Center for Gravitational Physics, Department of Space Science,
\\ Beihang University,  Beijing 100191, China}
\emailAdd{junkunzhao@buaa.edu.cn}
\abstract{We consider the effects of momentum relaxation on the topological quantum phase transitions in holographic Weyl semimetals. The translational symmetry breaking in the field theory is realized in the framework of massive gravity. We find that the critical value of the phase transition, characterized by the anomalous Hall conductivity, decreases with the increasing of graviton mass, i.e. the momentum relaxation strength. There exists a critical value of graviton mass above which the topological phase transition disappears and therefore the Weyl points are destroyed. All these phenomena are qualitatively similar to that of axion fields induced momentum relaxation, indicating that a universal feature emerges in the momentum relaxed holographic Weyl semimetals, which is also consistent with the predictions from weakly coupled field theory.}

\begin{document}
\maketitle
\flushbottom
\pagestyle{plain} \setcounter{page}{1}
\newcounter{bean}
\baselineskip16pt

%%%%%%%%%%%%% TITLEPAGE %%%%%%%%%%%%%%
%%%%%%%%%%%%%%%%%%%%%%%%%%%%%
\section{Introduction}
%%%%%%%%%%%%%%%%%%%%%%%%%%%%%
Weyl semimetals is a novel topological gapless state of quantum matter where its valence band and conduction band touch at certain points, namely the Weyl nodes, in momentum space \cite{Armitage:2017cjs,burkov0}. These Weyl nodes come in pairs with opposite chirality and are topologically stable under the perturbations that preserve charge conservation or translational symmetry \cite{Hosur:2013kxa}. Close to the Weyl nodes, the low-energy excitations satisfy the relativistic Weyl equation. Therefore, the quasiparticles behave like Weyl fermions which are anomalous quantum mechanically, leading to lots of exotic transport phenomena in Weyl semimetals. Thus, it has attracted numerous theoretical and experimental interest \cite{Armitage:2017cjs,burkov0,Hosur:2013kxa,Landsteiner:2016led,Dantas:2019rgp,Chernodub:2021nff}. Theoretically, most studies on Weyl semimetals are based on the topological band theory or the weakly coupled field theory. However, similar to graphene system \cite{jan}, the effective fine structure constant in Weyl semimetal can be large due to the smallness of the fermi velocity which plays the role of speed of light, implying that the Weyl semimetals can be strongly coupled without quasiparticles \cite{Gonzalez:2015tsa}. This raises the question of theoretical description of strong-interacting Weyl semimetals.

\para
On the other hand, gauge/gravity duality (or AdS/CFT correspondence) provides a novel approach to study the strongly-interacting quantum many-body system \cite{Zaanen:2015oix,book0,review}. The physics is holographically encoded in a weakly-coupled classical gravity living in one higher dimension. The holographic method has been successfully applied to explore various phases and their transports, yielding lots of significant insights. In the case of topological matter, the strongly-coupled holographic Weyl semimetals have been constructed recently \cite{Landsteiner:2015pdh,Landsteiner:2015lsa}, where the Weyl semimetal phase is characterized by a nonzero anomalous Hall conductivity and there exists a topological quantum phase transition between the Weyl semimetal phase and a topological trivial phase.\footnote{See \cite{Gursoy:2012ie,Jacobs:2015fiv,Hashimoto:2016ize,Fadafan:2020fod} for the semi-holographic and the top-down approach for strongly-coupled Weyl semimetals.} Subsequently, the existence of surface states \cite{Ammon:2016mwa} and the calculation of topological invariants \cite{Liu:2018djq} in the holographic system reveal the key features of Weyl semimetals. Many more works along this line can be found in \cite{Landsteiner:2016stv,Grignani:2016wyz,Copetti:2016ewq,Liu:2018bye,Ammon:2018wzb,Juricic:2020sgg,Baggioli:2018afg,
Liu:2018spp,Ji:2019pxx,Song:2019asj,Baggioli:2020cld,Liu:2020ymx,Zhao:2021pih,Ji:2021aan,Rodgers:2021azg} and see \cite{Landsteiner:2019kxb} for a recent review on the topic.

\para
In real condensed matter system, the translational symmetry is broken explicitly due to the presence of background lattice, which is important to relax the electron's momentum and give rise to a finite DC conductivities. Holographically, there are several different realizations of translational symmetry breaking, such as by the presence of a periodic lattice \cite{Horowitz:2012ky,Liu:2012tr} or by the mean field theory approaches (e.g. helical lattices \cite{Donos:2012js}, massive gravity \cite{Vegh:2013sk}, linear axion model \cite{Andrade:2013gsa}, Q-lattices \cite{Donos:2013eha}). Notably, the translational symmetry breaking has been missed in most studies of holographic Weyl semimetals.\footnote{See \cite{Ammon:2018wzb} for a study of quenched disorder on holographic Weyl semimetals in the probe limit.} This might be due to that holographic Weyl semimetal mainly focus on zero density physics and naively the momentum relaxation only play important role at finite density. However, in weakly coupled field theory descriptions, even at zero density the Weyl points can be annihilated by scattering with other Weyl points of opposite chirality due to the broken of translational symmetry \cite{Hosur:2013kxa}. Therefore, it is natural to study the role of translational symmetry breaking at strong coupling, i.e. in the holographic Weyl semimetals.

\para
Recently, the phenomenon of momentum relaxation in holographic Weyl semimetals has been explored in \cite{Zhao:2021pih}, where the presence of axion fields break the translational invariance \cite{Andrade:2013gsa,Baggioli:2021xuv}. It has been found that the Weyl semimetal phase shrinks and eventually ceases to exist with the increasing of momentum relaxation strength. However, there are two questions to be answered. The first question is the zero temperature ground-state of momentum relaxed holographic Weyl semimetals, which has distinct geometry structure because of the nonzero axion fields. It is important to construct the zero temperature solutions in order to reveal the underlying structure of the topological quantum phase transition. The second question is the universality of the effects of momentum relaxation on the system. In weakly coupled field theory \cite{Hosur:2013kxa}, the destroy of Weyl nodes should be independent of the mechanism of translational symmetry breaking. It is worth studying different model of translational symmetry breaking to test the universality of the phenomenon. In the present work, we will address the second question.

\para
We will use the massive gravity (see, eg. \cite{Vegh:2013sk,deRham:2010kj,Davison:2013jba,Blake:2013bqa,Cai:2014znn}) to test the effects of translational symmetry breaking on holographic Weyl semimetal. The bulk diffeomorphism invariance is broken due to the non-zero graviton mass terms, which corresponds to the translational symmetry breaking in the dual field theory. As the absolute zero temperature is not accessible in experiment, we will focus on the finite temperature physics. Even though, the information of quantum phase transition can be obtained due to the existence of quantum critical region. We will mainly focus on the effects of momentum relaxation on the properties of the topological phase transition in holographic Weyl semimetals.

\para
The outline of this paper is as follows. We begin, in section 2, by introducing the holographic model of Weyl semimetal with non-zero graviton mass terms. In section 3, we study the effects of graviton mass on d.c. transports of vector gauge field fluctuations. Section 4 is aimed to the conclusion and discussion. In the Appendix, we present the details of background equations of motion, asymptotic expansions and thermodynamics.

%%%%%%%%%%%%%%%%%%%%%%%%%%%%%
\section{Holographic setup }\label{sec2}
%%%%%%%%%%%%%%%%%%%%%%%%%%%%%
In this section, we introduce the holographic model of Weyl semimetals \cite{Landsteiner:2015pdh,Landsteiner:2015lsa} in the presence of non-zero graviton mass terms \cite{Vegh:2013sk,Cai:2014znn}. The action for the model is given by
\bea\label{action}
\mathcal{S}&=&\int d^5x\sqrt{-g}
\bigg[\frac{1}{2\kappa^2}\big(R+\frac{12}{L^2}\big)-\frac{1}{4}F_{ab}F^{ab}-\frac{1}{4}\mathcal{F}_{ab}\mathcal{F}^{ab}
+\frac{\alpha}{3}\epsilon^{abcde}A_a\big(F_{bc}F_{de} +3\mathcal{F}_{bc}\mathcal{F}_{de}\big) \nn\\
&&-(D_a\Phi)^\ast(D^a\Phi)-V(\Phi)+\frac{m_g^2}{2\kappa^2}\sum_{i=1}^{4}c_i \mathcal{U}_i(g,f) \bigg]+\mathcal{S}_{\text{GH}}+\mathcal{S}_{\text{c.t.}} \,,
\eea
where $\kappa^2$, $L$ and $\alpha$ are the gravitational constant, AdS radius and Chern-Simons coupling respectively. The axial gauge field $A_\mu$ is dual to the axial current in the field theory and its field strength is $F_{ab}=\partial_a A_b-\partial_b A_a$. The vector gauge field $V_\mu$ is dual to the vector current in the field theory and its field strength is $\mathcal{F}_{ab}=\partial_a V_b-\partial_b V_a$. The Chern-Simons terms are included to characterize the anomalous of the axial symmetry in the field theory. The complex scalar field $\Phi$ is axial charged with the covariant derivative $D_a=\partial_a-i q A_a$. We choose the potential of the scalar field as $V(\Phi)=m^2\Phi^2+\frac{\lambda}{2}\Phi^4$ with the scalar field mass $m^2=-3$. The graviton mass terms are linear combination of $\mathcal{U}_i$, where the coefficient $c_i$ are dimensionless constants and $m_g$ is the graviton mass. Note that $\mathcal{U}_i$ are symmetric polynomials of the eigenvalues of the $5\times5$ matrix $\mathcal{K}^a_{\ b}\equiv \sqrt{g^{ac}f_{cb}}$
\bea
\mathcal{U}_1&=&[\mathcal{K}] \,, \nn\\
\mathcal{U}_2&=&[\mathcal{K}]^2-[\mathcal{K}^2] \,, \nn\\
\mathcal{U}_3&=&[\mathcal{K}]^3-3[\mathcal{K}][\mathcal{K}^2]+2[\mathcal{K}^3] \,, \nn\\
\mathcal{U}_4&=&[\mathcal{K}]^4-6[\mathcal{K}^2][\mathcal{K}]^2+8[\mathcal{K}^3][\mathcal{K}]+3[\mathcal{K}^2]^2-6[\mathcal{K}^4] \,,\nn
\eea
where the rectangular brackets denote traces: $[\mathcal{K}]=\mathcal{K}^a_{\ a}$. In massive gravity, the dynamical metric $g_{ab}$ couples to the symmetric reference metric $f_{cd}$, which breaks diffeomorphism invariance and gives the graviton a mass. We choose the reference metric $f_{cd}=\text{diag}(0, F, F, F_z, 0)$ with constant $F$ and $F_z$. Thus the spatial reparameterization symmetry is broken, which results in the momentum relaxation in the field theory. $\mathcal{S}_{\text{GH} }$ is the standard Gibbons-Hawking term and $\mathcal{S}_{\text{c.t.} }$ is the counterterm to make the physical observable finite. For simplicity, we will concentrate on the case of $q=1$ and $\lambda=1/10$ in this paper.

\para
We set $2\kappa^2=1$. The bulk equations of motion are
\bea
R_{ab}-\frac{1}{2}g_{ab}\big(R+12\big)-m_g^2\chi_{ab}-\frac{1}{2}T_{ab}&=&0 \,,\nn\\
\nabla_b \mathcal{F}^{ba}+2\alpha\epsilon^{abcde}F_{bc}\mathcal{F}_{de}&=&0\,,\nn\\
\nabla_b F^{ba}+\alpha\epsilon^{abcde}(F_{bc}F_{de}+\mathcal{F}_{bc}\mathcal{F}_{de})
   +i q[\Phi(D^a\Phi)^\ast-\Phi^\ast(D^a\Phi) ]&=&0\,,\nn\\
D_aD^a\Phi-m^2\Phi-\lambda (\Phi^\ast)^2 \Phi &=&0 \,. \nn
\eea
where
\bea
\chi_{ab}&=&\frac{c_1}{2} \big( \mathcal{U}_1 g_{ab}-\mathcal{K}_{ab} \big)
      +\frac{c_2}{2} \big( \mathcal{U}_2g_{ab}-2\mathcal{U}_1\mathcal{K}_{ab}+2\mathcal{K}^2_{ab} \big) \nn\\
   &&+\frac{c_3}{2} \big( \mathcal{U}_3g_{ab}-3\mathcal{U}_2\mathcal{K}_{ab}+
      6\mathcal{U}_1\mathcal{K}^2_{ab}-6\mathcal{K}^3_{ab} \big) \nn\\
   &&+\frac{c_4}{2} \big( \mathcal{U}_4g_{ab}-4\mathcal{U}_3\mathcal{K}_{ab}+
     12\mathcal{U}_2\mathcal{K}^2_{ab}-24\mathcal{U}_1\mathcal{K}^3_{ab}+24\mathcal{K}^4_{ab} \big) \,,\nn\\
T_{ab}&=&\bigg[\mathcal{F}_{ac}\mathcal{F}_{b}^{~c}-\frac{1}{4}g_{ab}\mathcal{F}^2\bigg]
    +\bigg[F_{ac}F_{b}^{~c}-\frac{1}{4}g_{ab}F^2\bigg] \nn\\
    &&+\big[  D_a\Phi (D_b\Phi)^\ast+(D_a\Phi)^\ast D_b\Phi \big] -g_{ab} \bigg[ (D_c\Phi)^\ast(D^c\Phi)+V(\Phi) \bigg] \,, \nn
\eea

\para
We make the following ansatz for the background fields
\bea\label{ansatz}
ds^2=-udt^2+\frac{dr^2}{u}+f(dx^2+dy^2)+hdz^2,\,\,  A=A_z dz,\,\,  \Phi=\phi(r),
\eea
where $u, f, h, A_z, \phi$ only depend on the radial coordinate $r$. With the above ansatz, we find
\bea
&& \mathcal{K}_{ab}=\text{diag}  \Big( 0 \,\,, F\sqrt{f} \,\,, F\sqrt{f} \,\,, F_z\sqrt{h} \,\,, 0 \Big)  \,,\\
&& \mathcal{U}_1=\frac{2F}{\sqrt{f} }+\frac{F_z}{\sqrt{h} } \,,\quad  \mathcal{U}_2=\frac{2F^2}{f}+\frac{4FF_z}{\sqrt{fh} } \,,\quad  \mathcal{U}_3=\frac{6F^2F_z}{f\sqrt{h} }  \,,\quad  \mathcal{U}_4=0 \,.
\eea
We can obtain the corresponding equations of motion following the above ansatz and see appendix \ref{app:a} for details. As $r\to \infty$, the background geometry is asymptotically to $AdS_5$ with $u, f, h \sim r^2+\cdots$. For the scalar field and the axial gauge field, we have
\bea
\phi=\frac{M}{r}+\cdots ,\quad\quad  A_z=b+\cdots ,
\eea
where $M$ and $b$ correspond to the mass parameter and the time-reversal symmetry breaking parameter of the field theory respectively.

%%%%%%%%%%%%%%%%%%%%%%%%%%%%%
\subsection{Brief review of holographic Weyl semimetal}
%%%%%%%%%%%%%%%%%%%%%%%%%%%%%
Before discussing the effects of momentum relaxation induced by the massive gravity, it is useful to summarise the important ingredients in translation invariant holographic Weyl semimetals \cite{Landsteiner:2015lsa,Landsteiner:2015pdh}. In this subsection, we will review the zero and the finite temperature physics of the holographic Weyl semimetals and focus on their properties of phase transition.

\begin{figure}[!ptb]
\begin{center}
\includegraphics[width=0.57\textwidth]{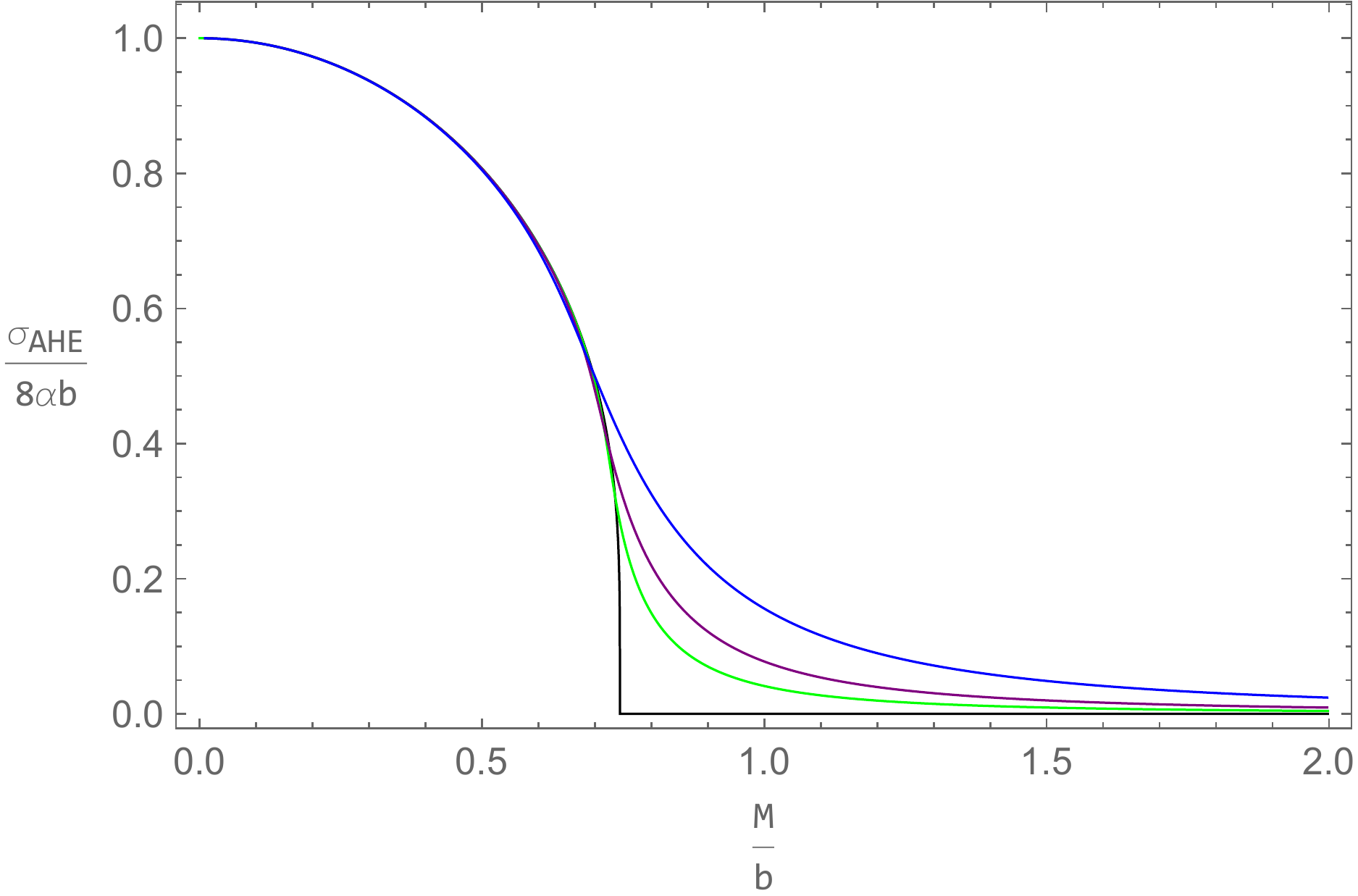}
\end{center}
\vspace{-0.6cm}
\caption{The anomalous Hall conductivity as a function of the $M/b$ in the minimal holographic Weyl semimetals \cite{Landsteiner:2015pdh}. The black line corresponds to the zero temperature while the colored lines correspond to the finite temperature with $T/b=0.05$ (blue), $0.03$ (purple), $0.02$ (green) respectively. At finite temperature, the sharp quantum phase transition becomes a crossover.}
\label{fig:ahe0}
\end{figure}

\para
At zero temperature, we have $u=f$ which corresponds to Lorentz invariance in the field theory along the $(t, x, y)$ direction. We have only one tunable dimensionless parameter $M/b$. By tuning $M/b$, we can find three phases with different infrared solutions \cite{Landsteiner:2015pdh}: (I) {\em the Weyl sememetal phase} exists when $M/b<(M/b)_c$, (II) {\em the Lifshitz critical point} exists when $M/b=(M/b)_c=0.744$, (III) {\em the topological trivial phase} exists when $M/b>(M/b)_c$. These three phases are distinguished by anomalous Hall conductivity, which is nonzero in the Weyl semimetal phase while vanishes in the topological trivial phase and the critical point. With the increasing of $M/b$, the system experiences a topological quantum phase transition from the Weyl semimetal phase to a topological trivial phase and see Fig.\ref{fig:ahe0} for the phase diagram. Note that, the anomalous Hall conductivity is proportional to the near horizon value of the axial gauge field, i.e. $\sigma_{AHE}\propto A_z(0)$, which is the order parameter of the quantum phase transition.
%%%%%%%%%%%%%%%%%%%%%%%%%%%%
%%%%%%%%%%%%%%%%%%%%%%%%%%%%%
\para
At finite temperature, the infrared (IR) fixed points are covered by black hole horizon, where the background fields admit a regular expansion near the horizon. The topological quantum phase transition now becomes a smooth crossover due to thermal correlations. In particular, the anomalous Hall conductivity has a very small value in the topological trivial phase and the quantum critical point produces a quantum critical range at finite temperature. Even though, we can still obtain the critical point of phase transition from the behavior of anomalous Hall conductivity. We define the point $M/b$ with maximal $|\frac{ \partial \sigma_{AHE} }{ \partial (M/b) }|$ as the critical value of the phase transition, which will approach to the critical point of topological quantum phase transition as the temperature is decreased. For example, the critical value at $T/b=0.02$ is $0.722$ with a relative error within $3\%$ compared with the zero temperature result. Therefore, we will use this method to identify the critical point of the phase transition in the following study.

%%%%%%%%%%%%%%%%%%%%%%%%%%%%%
\subsection{Holographic Weyl semimetal with massive gravity}
%%%%%%%%%%%%%%%%%%%%%%%%%%%%%
In this subsection, we will obtain the equilibrium solutions with non-zero graviton mass terms by solving the background equations of motion. It is reasonable to postulate that there still exists a quantum phase transition from the Weyl semimetal phase to a topological trivial phase. We will focus on the finite temperature physics of the momentum relaxed holographic Weyl semimetals.

\para
Before studying the finite temperature physics, let us point out one important fact. From the $tt$ and $rr$ component of Einstein equation, we know \footnote{This is the first equation of the full background equations of motion in the appendix \ref{app:a}.}
\bea\label{eqn:uf}
\big[\sqrt{h} (u'f-u f')\big]'=-m_g^2\big[6F^2F_zc_3+2FF_z c_2\sqrt{f}+2F^2 c_2\sqrt{h}+F c_1\sqrt{f h} \big].
\eea
In the translation invariant case, i.e. $m_g=0$, the right hand side is zero. The above equation saturates the null energy condition (NEC), which requires $\big[\sqrt{h}(u'f-uf') \big]'\geq0$. From this equation, we can also define a radially conserved Noether charge $Q=\sqrt{h} (u'f-uf')$, which gives $Q=Ts$ at the black hole horizon. Consequently, the zero temperature solution (i.e. for $u=f$) is equivalent to $Q=0$. However, in the presence of non-zero graviton mass terms, we can not convert (\ref{eqn:uf}) into a total derivative and the $u=f$ condition for ground-state is not applicable. The situation here is similar to the case in Ref.\cite{Zhao:2021pih}, indicating that the zero temperature ground-state has different geometries with $u\neq f$. Therefore, the study of ground state demands more works and we will leave it for further study.

\para
Additionally, we can restrict the value of $c_i, (i=1,2,3)$ from (\ref{eqn:uf}). The observation is that the right hand side of (\ref{eqn:uf}) should be positive in order to maintain the NEC in the $m_g\to 0$ limit. In the case where only one of $c_i$ is nonzero, this restriction requires $c_i<0$.\footnote{It is possible to have positive $c_i$ in the parameter space by considering more nonzero $c_i$.}

%%%%%%%%%%%%%%%%%%%%%%%%%%%%\textwidth
\begin{figure}[!ptb]
\begin{center}
\includegraphics[width=0.45 \textwidth]{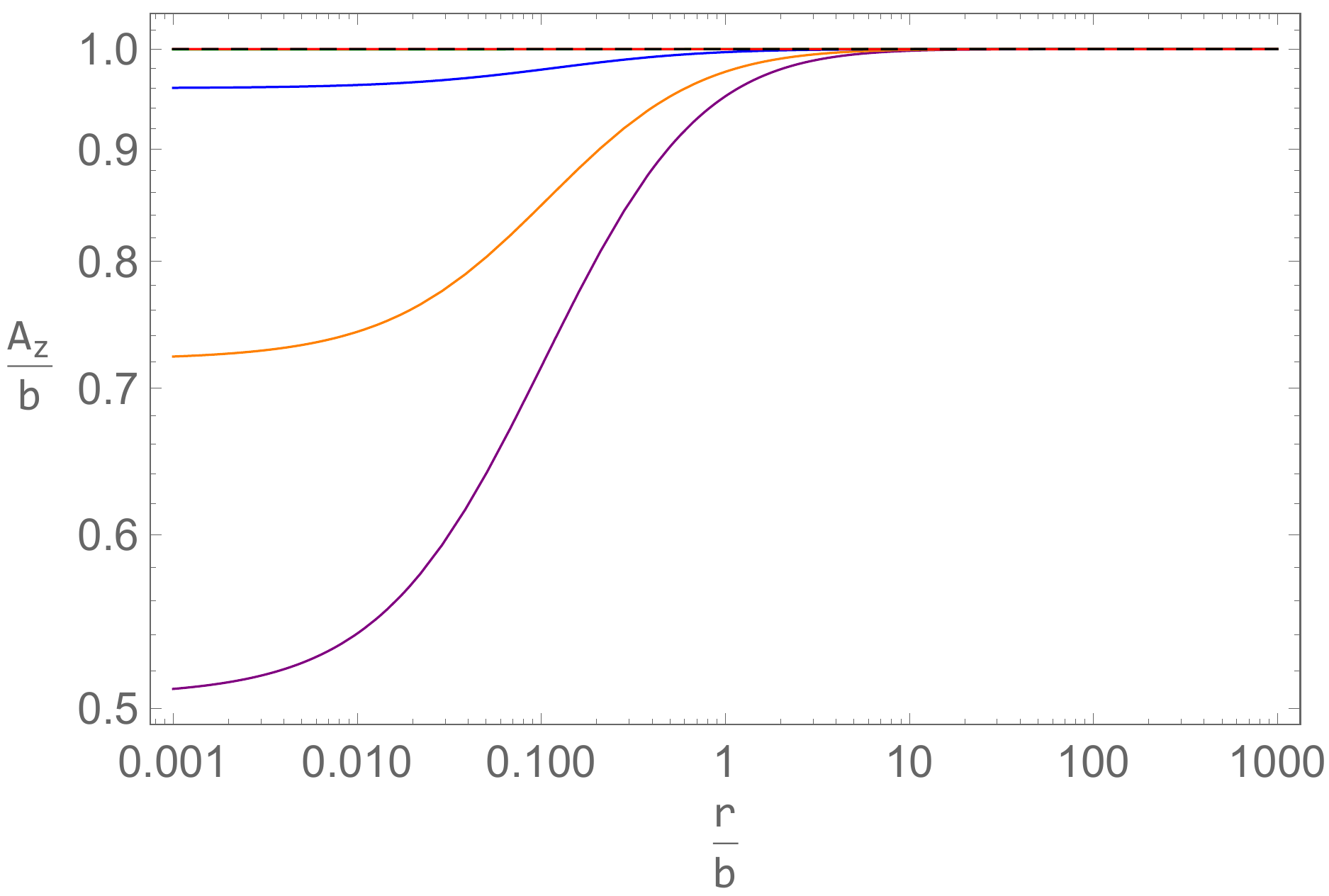}
\includegraphics[width=0.45 \textwidth]{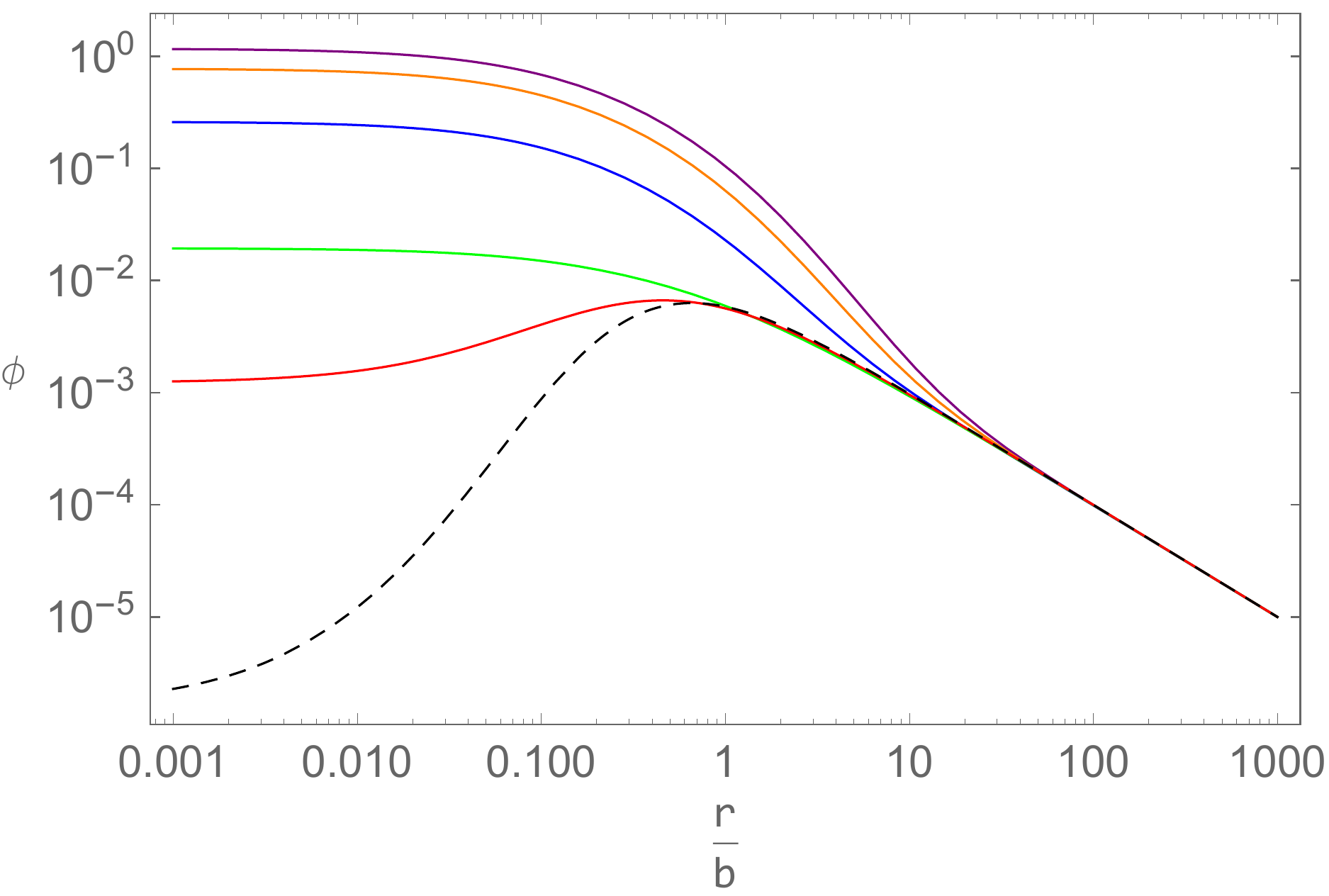}
\end{center}
\vspace{-0.6cm}
\caption{Log-log plot of the bulk profile of $A_z$ (left) and $\phi$ (right) for different graviton mass $m_g|c_2|/b$ at $M/b=0.01$. As $m_g|c_2|/b$ is increased from $0$ to $7$, the gauge field $A_z$ changes from top to bottom while the scalar field $\phi$ changes from bottom to top.}
\label{fig:bg}
\end{figure}
%%%%%%%%%%%%%%%%%%%%%%%%%%%%%

\para
Now, we begin our study of the finite temperature physics, where the system is specified by five dimensionless parameters $M/b, T/b$ and $m_g|c_i|/b, (i=1,2,3)$. For simplicity, we will focus on the $\mathcal{U}_2$ sector of massive gravity by tuning the parameter $c_2$ and fixing $m_g=1$. We further fix the temperature $T/b=0.03$ and the reference metric $F=F_z=1$. Therefore, the system now has two dimensionless parameters $M/b$ and $m_g|c_2|/b$. The infrared (IR) and ultraviolet (UV) expansions for the background fields can be found in the appendix \ref{app:a}. For given initial seeds, the background solutions can be obtained by integrating the equations of motion from IR towards UV.

\para
Specifically, in the Weyl semimetal phase, we find that the background fields flow differently from UV to IR for different graviton mass. This is illustrated in Fig.\ref{fig:bg}, where we show the profile of $\phi$ and $A_z$ for different graviton mass at $M/b=0.01$. In the translation invariant case (i.e. $m_g=0$, the black dashed lines), the axial gauge field keeps a constant value, while the scalar field first increases, then decreases to zero from the boundary towards the horizon, showing a non-monotonic behavior. In the non-zero graviton mass case, however, the axial gauge field turns out to decrease monotonically from the boundary to the horizon. The scalar field shows a transition from a non-monotonic function to a monotonic increasing function from the boundary towards the horizon. Especially, as the graviton mass is increased, the near horizon value of $A_z$ decreases while the near horizon value of $\phi$ increases. Therefore, the presence of non-zero graviton mass terms can lead to a dramatic change of the flow of matter fields. This signals that an emergent phenomenon occurs in the Weyl semimetal phase under momentum relaxation, which will be confirmed from the critical point of the phase transition studied in the next section.

%%%%%%%%%%%%%%%%%%%%%%%%%%%%%
\section{Effects of massive gravity on phase transition}\label{sec3}
%%%%%%%%%%%%%%%%%%%%%%%%%%%%%
In this section, we will explore the effects of graviton mass on the transport properties of the holographic Weyl semimetals. We will compute the anomalous Hall conductivity and obtain the phase diagram of the system. We will discuss the spontaneous symmetry breaking solution at $M/b=0$ for large graviton mass. We will also study the longitudinal and transverse conductivity.

\para
Using the Kubo formula, the conductivities of the dual field theory reads
\bea
\sigma_{ij}=\lim_{\omega\to 0}\frac{1}{i\omega}\langle J_i J_j\rangle_R(\omega, \mathbf{k}=0) \,,
\eea
where the current-current retarded Green's functions can be computed by studying the bulk gauge field fluctuations with ingoing boundary condition at the horizon.

\para
The vector gauge field perturbations take the form
\bea\label{equ:perb}
\delta V_x=v_x(r)e^{-i\omega t},\,\,  \delta V_y=v_y(r)e^{-i\omega t},\,\,  \delta V_z=v_z(r)e^{-i\omega t} \,,
\eea
Note that, the vector gauge field fluctuations decouple from the metric fluctuations. Therefore, the situation here is different with the finite density cases in \cite{Vegh:2013sk,Davison:2013jba,Blake:2013bqa}. For finite density system, the non-zero graviton mass terms usually affect the physics of the system in two ways \cite{Vegh:2013sk}. First, in thermal equilibrium, the background geometry is altered by the non-zero mass terms and especially the near horizon geometry has an $\text{AdS}_2\times \mathbb{R}^d$ structure at zero temperature. Second, at the level of linearised perturbations, the dynamical degree of freedoms are increased and the dual energy momentum tensor is not conserved due to the non-zero graviton mass terms. In contrast, as the holographic Weyl semimetals is a zero density system, the decoupling of vector gauge field fluctuations with the metric fluctuations indicates that the effects of non-zero mass terms on the transports arise through their effects on the equilibrium solutions.\footnote{Instead, the axial gauge field fluctuations will necessary involve the metric fluctuations. It will be interesting to invest the axial conductivity in the presence of non-zero mass terms and verify the $1/3$ relation between axial Hall conductivity and electric Hall conductivity found in \cite{Copetti:2016ewq}.}

\para
Plugging (\ref{equ:perb}) into the vector gauge field equation, we obtain
\bea
v_z''+\bigg(\frac{u'}{u}+\frac{f'}{f}-\frac{h'}{2h}\bigg)v_z'+\frac{\omega^2}{u^2}v_z&=&0  \,,\\
v_\pm''+\bigg(\frac{u'}{u}+\frac{h'}{2h}\bigg)v_\pm'+\frac{\omega^2}{u^2}v_\pm \pm 8\alpha\omega\frac{A_z'}{u\sqrt{h}}v_\pm&=&0  \,,
\eea
where $v_\pm=v_x\pm i v_y$. Since we are interested in the DC conductivities, we will use the near-far matching method \cite{Landsteiner:2015pdh} to compute these quantities. Finally, the DC conductivities $\sigma_{xx}, \sigma_{yy}$ and $\sigma_{xy}$ reads
\bea\label{eq:ahe}
\sigma_T=\sigma_{xx}=\sigma_{yy}=\frac{G_{+}+G_{-}}{2i\omega}=\sqrt{h(r_h)},\,\,  \sigma_{xy}=\frac{ G_{+}-G_{-} }{2\omega}=8\alpha \big( b-A_z(r_h) \big) \,,
\eea
where $G_{\pm}=\omega \big( \pm 8\alpha(b-A_z(r_h))+i \sqrt{h(r_h)} \big)$ is the Green functions of $v_\pm$. Similarly, the longitudinal conductivity $\sigma_{zz}$ is given by
\bea
\sigma_{zz}=\frac{G_{zz}}{i\omega}=\frac{f(r_h)}{\sqrt{h(r_h)}} \,,
\eea
The anomalous Hall conductivity for the consistent current reads
\bea\label{equ:ahe}
\sigma_{\text{AHE}}=8\alpha b-\sigma_{xy}=8\alpha A_z(r_h) \,,
\eea
%%%%%%%%%%%%%%%%%%%%%%%%%%%%%
\subsection{Anomalous Hall conductivity}\label{sec:ahe}
\begin{figure}[!ptb]
\begin{center}
\includegraphics[width=0.57\textwidth]{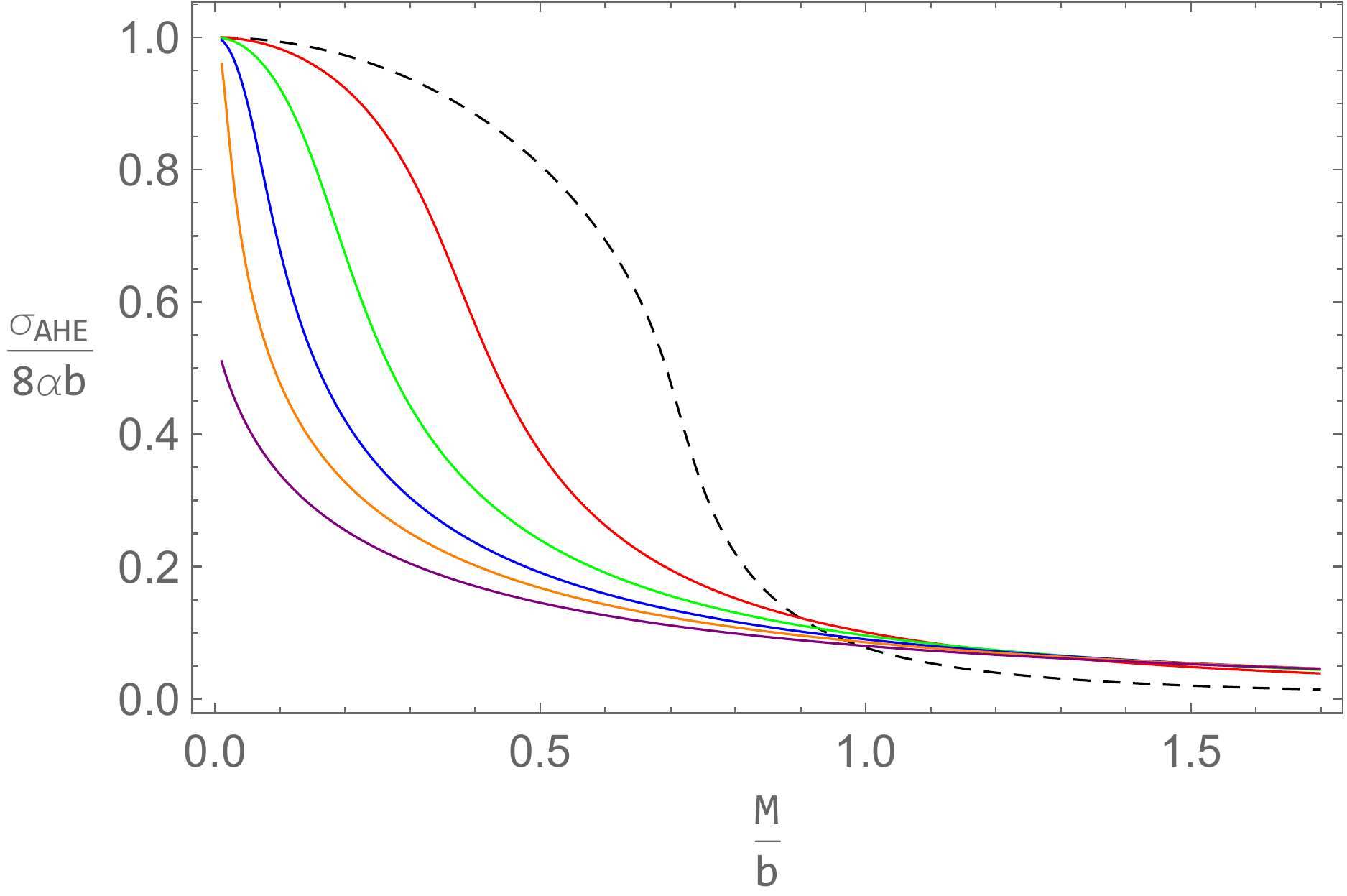}
\end{center}
\vspace{-0.6cm}
\caption{The anomalous Hall conductivity as a function of $M/b$ for different graviton mass $m_g|c_2|/b$ at $T/b=0.03$. The black dashed line is for the massless case, while the colored solid line are for graviton mass $m_g|c_2|/b=0.5$ (red), $1.5$ (green), $3$ (blue), $4.5$ (orange), $7$ (purple) from right to left respectively.}
\label{fig:ahec2}
\end{figure}

\begin{figure}[!ptb]
\begin{center}
\includegraphics[width=0.57\textwidth]{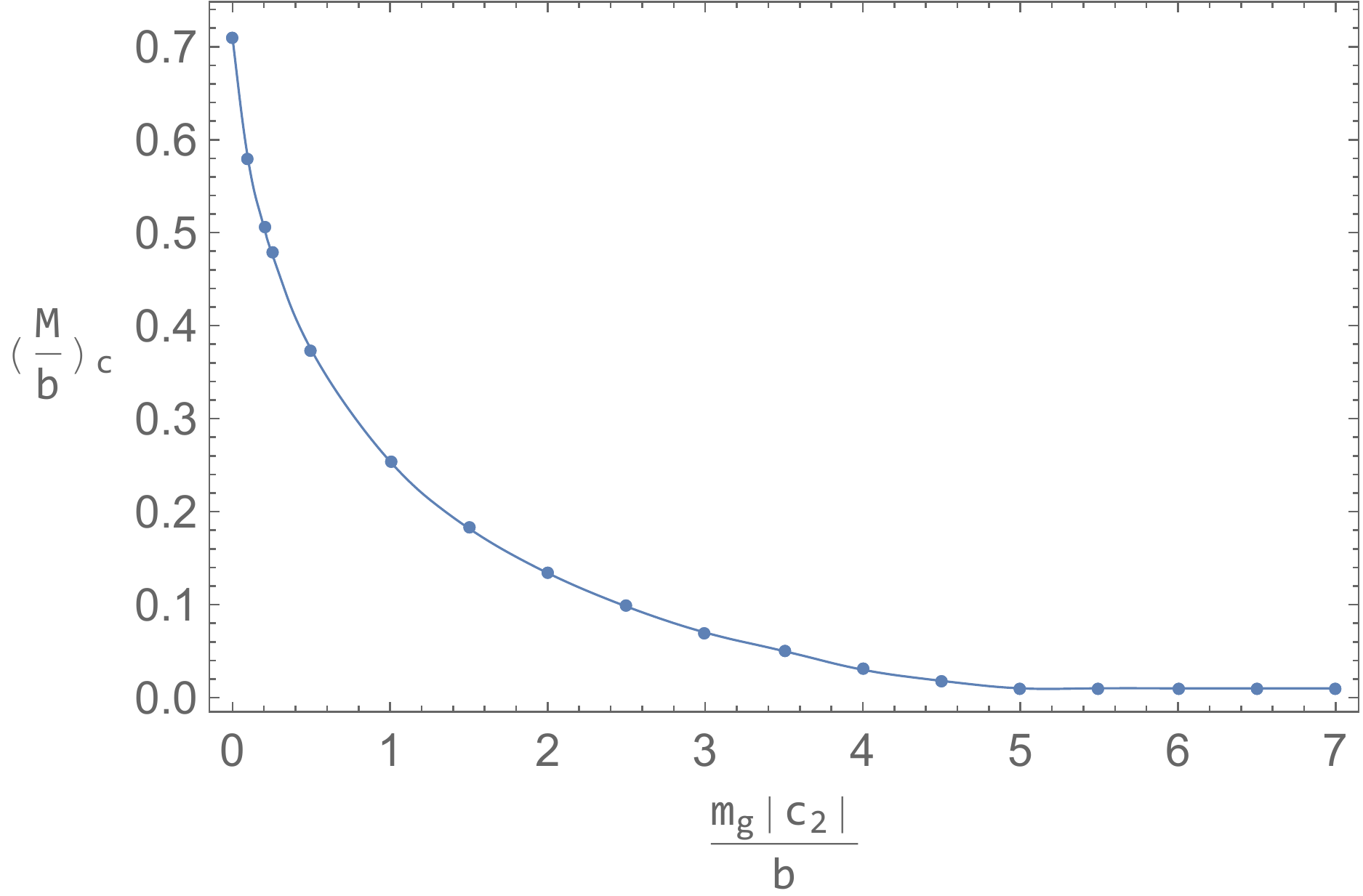}
\end{center}
\vspace{-0.6cm}
\caption{The critical point as a function of graviton mass $m_g|c_2|/b$ at $T/b=0.03$. }
\label{fig:critical}
\end{figure}

%%%%%%%%%%%%%%%%%%%%%%%%%%%%%
We will now study the anomalous Hall conductivity. Fig.\ref{fig:ahec2} shows the anomalous Hall conductivity as a function of $M/b$ at $T/b=0.03$, where each curve corresponds to different graviton mass $m_g|c_2|/b$. For the translation invariant case (zero graviton mass), as $M/b$ is increased, the (normalized) anomalous Hall conductivity decreases monotonically from $1$ in the Weyl semimetal phase to a very small value in the topological trivial phase. For a non-zero graviton mass, we find a similar monotonically decreasing behavior with the increasing of $M/b$. The anomalous Hall conductivity decreases rapidly and becomes negligible at large value of $M/b$. As the graviton mass becomes large, the anomalous Hall conductivity still decrease monotonically. However, near $M/b\approx0$, its value is less than $1$ and decreases with the increasing of graviton mass.\footnote{We will explain this point in the following subsection.}
%%%%%%%%%%%%%%%%%%%%%%%%%%%%%
%%%%%%%%%%%%%%%%%%%%%%%%%%%%%
\para
To characterise the effects of non-zero graviton mass terms on the properties of phase transition, we will also study the critical point $(M/b)_c$ as a function of graviton mass. The critical value of the phase transition is encoded in the anomalous Hall conductivity, which is the point with maximum $|\frac{ \partial \sigma_{AHE} }{ \partial (M/b) }|$. Fig.\ref{fig:critical} shows that as the graviton mass is increased, the critical value $(M/b)_c$ decreases monotonically. Above a critical graviton mass with $m_g|c_2|/b\approx5$, the critical value $(M/b)_c$ goes to zero. Note that, the Weyl semimetal phase is characterized by the non-zero anomalous Hall conductivity and exists for $M/b<(M/b)_c$. Therefore, the behavior of critical point shows that the non-zero graviton mass terms can destroy the Weyl points and reduce the region of Weyl semimetal phase.
%%%%%%%%%%%%%%%%%%%%%%%%%%%%%
%%%%%%%%%%%%%%%%%%%%%%%%%%%%%
\para
The phenomenon described above are similar to the axion fields induced momentum relaxation studied in Ref.\cite{Zhao:2021pih}. Note that, the massive gravity studied in this paper differ from the linear axion model, as their background equations of motion are different. Therefore, regardless of the different mechanism of translational symmetry breaking in two models, their effects of momentum relaxation on the holographic Weyl semimetals are universal.

\para
All these phenomena are also consistent with the predictions from weakly coupled field theory \cite{Hosur:2013kxa}. We will now give an intuitive picture, similar to the Ref.\cite{Zhao:2021pih}, to explain it.\footnote{There are opinions that the massive gravity captures the phenomenon of disorder in field theory. One may interpret the phenomenon here as disorder-induced localization for the topological degrees of freedom. We thank Francisco Pena-Benitez for pointing out this.} The explanation starts by relating $k_L$ (the width of Brillouin zone) to the graviton mass $m_g|c_2|/b$, and we fix the distance between Weyl points to be $1$. Thus, it is natural to expect that $k_L$ will decrease with the increasing of graviton mass, as the massless case corresponds to translation invariance with $k_L\to\infty$. Therefore, there exists a critical $m_g|c_2|/b$ to make $k_L=1$ where the two Weyl points meet at the boundary of Brillouin zone and annihilate with each other. The disappearance of the Weyl semimetal phase can be understood as a result of the annihilation of Weyl points.

%%%%%%%%%%%%%%%%%%%%%%%%%%%%%
\subsection{Spontaneous symmetry breaking solutions at $M/b=0$}\label{sec:ssb}
%%%%%%%%%%%%%%%%%%%%%%%%%%%%%
To explain the behavior of anomalous Hall conductivity near $M/b\approx0$, we will discuss two solutions of the background system at $M/b=0$. First, we have an analytical solution \cite{Cai:2014znn}
\bea\label{exact}
u&=&r^2-\frac{r_h^4}{r^2}+\frac{Fm_g^2c_1}{3}r(1-\frac{r_h^3}{r^3})+F^2m_g^2c_2(1-\frac{r_h^2}{r^2})
  +\frac{2F^3m_g^2c_3}{r}(1-\frac{r_h}{r}), \\
f&=&h=r^2, \quad\quad\quad\quad\quad   A_z=b, \quad\quad\quad\quad\quad   \phi=0, \nn
\eea
Note that, this solution is equal to the analytical solution in Ref.\cite{Zhao:2021pih}, if we choose $c_2=-\frac{\beta^2}{4F^2m_g^2}$ and $ c_1=c_3=0$. \footnote{The equality between a sector of massive gravity and the axion model is first shown in\cite{Andrade:2013gsa}. Similarly, the solution of massive gravity can also be mapped to that of conformal gravity in 4d \cite{EslamPanah:2019fci}.} From (\ref{equ:ahe}), we obtain $\frac{\sigma_{AHE} }{8\alpha b}|_{M=0}=1$, which is different with the numerical results of anomalous Hall conductivity for large graviton mass near $M/b\approx0$.

\para
Second, in addition to the analytical solution (\ref{exact}), the system also has a spontaneous symmetry breaking (SSB) solution. This comes from the observation that the zero temperature near-horizon limit of (\ref{exact}) is $\text{AdS}_2\times \mathbb{R}^3 $. At zero temperature, the near horizon expansion of metric function takes $u=u_2(r-r_h)^2+\cdots =(2-c_2 F^2 m_g^2/r_h^2- 2c_3F^3 m_g^2/r_h^3) (r-r_h)^2+\cdots $. By analyzing the scalar field fluctuations following \cite{Horowitz:2009ij}, its effective mass near the $\text{AdS}_2$ horizon reads $m^2_{eff}=\frac{m^2}{u_2}+\frac{b^2 q^2}{r_h^2u_2}$. Thus, if the effective mass is below the Breitenlohner-Freedman (BF) bound of $\text{AdS}_2$, i.e. $m^2_{eff}<m^2_{BF}=-1/4$, the scalar field perturbations are unstable. This instability indicates the existence of new phases with non-trivial $\phi$, which can be identified as a spontaneous symmetry breaking solution. The condition for instability can be computed analytically at zero temperature, which is determined by $m_g|c_i|/b$. As we turn on the temperature and keep $T/b$ fixed, there exists a critical $m_g|c_i|/b$ above which the SSB solution appears at $M/b=0$. We show the condition for instability in the table for $m_g|c_2|/b$.
%%%%%%%%%%%%%%%%%%%%%%%%%%%%%
\begin{table}[h]
\begin{center}\label{tab:1}
\begin{tabular}{|c|c|c|}
\hline
                   & $ T/b=0 $    &   $ T/b=0.03 $   \\
\hline\hline
$m_g |c_2|/b $     & $   1   $    &   $   5.45   $   \\
\hline
\end{tabular}
\end{center}
\vspace{-0.4cm}
\caption{\small The instability condition for scalar field perturbations.}
\end{table}
%%%%%%%%%%%%%%%%%%%%%%%%%%%%%
\para
In summary, there exist two branches of background solutions at $M/b=0$ for large graviton mass. When $m_g|c_2|/b$ is bigger than the critical value shown in the table, the background solutions approach to the spontaneous symmetry breaking solution in the $M/b\to 0$ limit. Therefore, its anomalous Hall conductivity will deviate $1$ and generally depend on $m_g|c_2|/b$, as is shown in Fig.\ref{fig:ahec2} for $M/b\to0$. However, a detailed analysis of the physics near $M/b=0$ is beyond the scope of this paper and demands more works.

%%%%%%%%%%%%%%%%%%%%%%%%%%%%%
\subsection{Transverse and longitudinal conductivities}
\begin{figure}[!ptb]
\begin{center}
\includegraphics[width=0.45\textwidth]{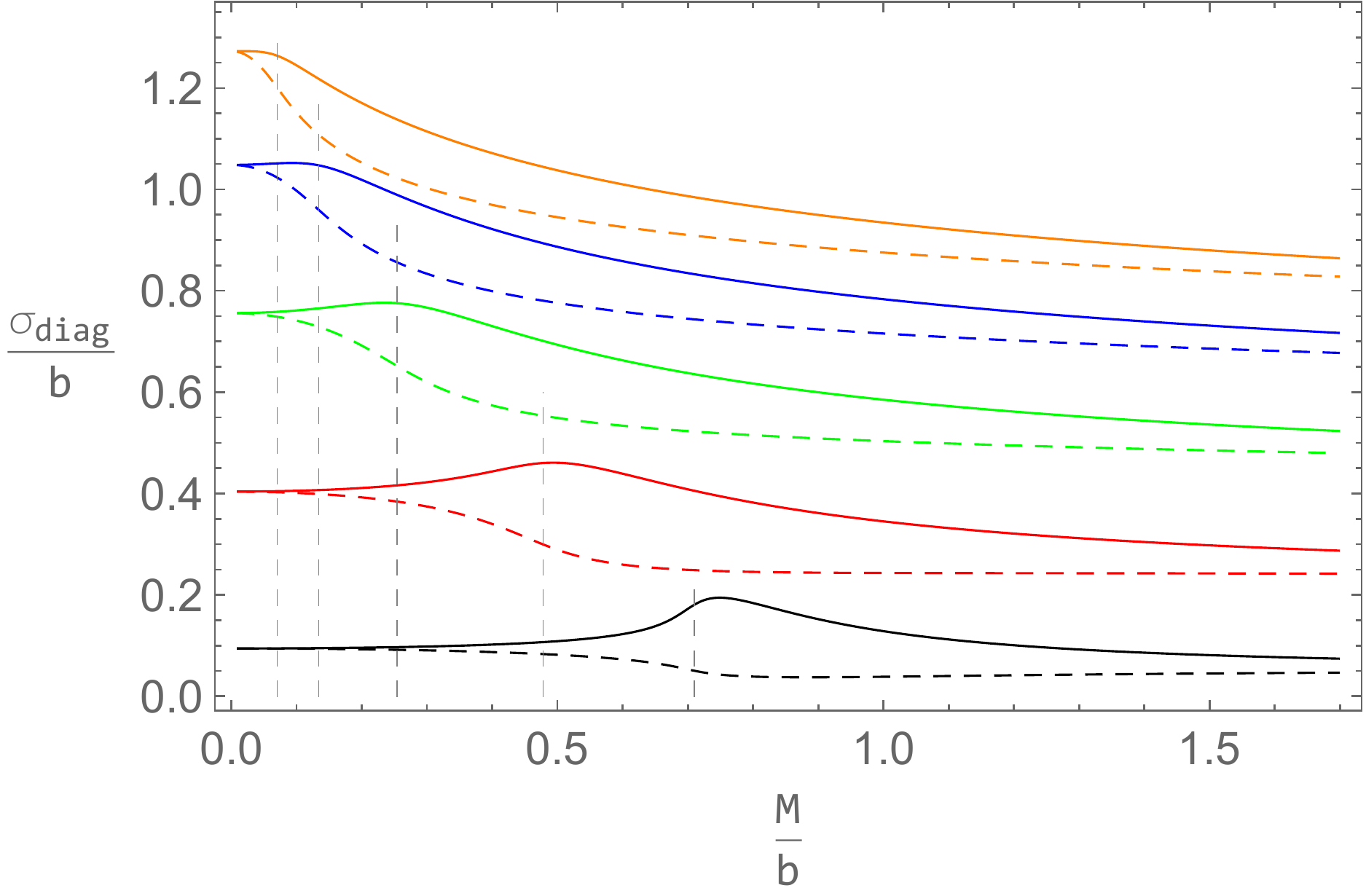}
\includegraphics[width=0.45\textwidth]{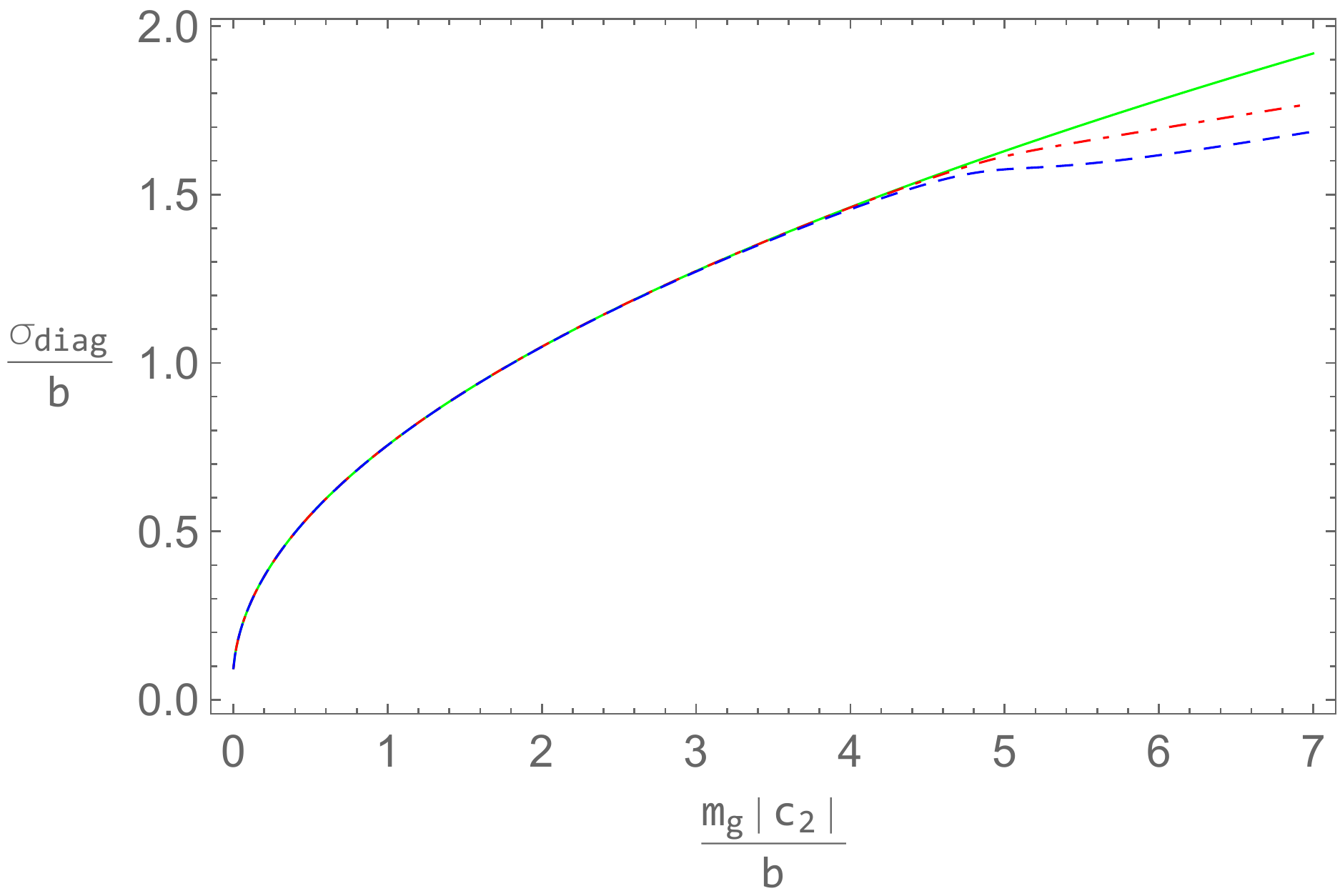}
\end{center}
\vspace{-0.6cm}
\caption{Left: The transverse (solid lines) and longitudinal (dashed lines) conductivities as a function of $M/b$ for different graviton mass at $T/b=0.03$. The plot is for $m_g|c_2|/b=0$ (black), $0.25$ (red), $1$ (green), $2$ (blue), $3$ (orange). The dashed gray lines are the location of critical points of the phase transition. Right: The diagonal conductivities as a function of graviton mass. The green line is for the analytical result at $M/b=0$, while the red (blue) dashed lines are for numerical results of transverse (longitudinal) conductivities at $M/b$=0.01. }
\label{fig:diag}
\end{figure}

{\label{sec:diag}}
%%%%%%%%%%%%%%%%%%%%%%%%%%%%%
Apart from the anomalous Hall conductivity, it is interesting to study the transverse ($\sigma_T$) and longitudinal ($\sigma_L$) conductivities as a function of $M/b$ for different graviton mass, which is shown in the left plot Fig.\ref{fig:diag}. For fixed graviton mass, the $\sigma_T$ and $\sigma_L$ have a same value at $M/b=0.01$. Then with the increasing of $M/b$, the $\sigma_T$ increase while the $\sigma_L$ decrease. At the intermediate range of $M/b$, the $\sigma_T$ produces a peak while the $\sigma_L$ produces a minimal. Finally, they both approach to a constant value at large $M/b$. On the other hand, as we increase the graviton mass, the diagonal conductivities are increased. Especially, the peak and minimal of the diagonal conductivities move to a small value of $M/b$ with the increasing of graviton mass, which is similar to the behavior of critical point of the phase transition.
%%%%%%%%%%%%%%%%%%%%%%%%%%%%%
%%%%%%%%%%%%%%%%%%%%%%%%%%%%%

\para
At $M/b=0.01$, the $\sigma_T$ and $\sigma_L$ are equal, which can be compared to the conductivities obtained from the analytical solution (\ref{exact}) (i.e. $\sigma_{\text{diag}}=\frac{1}{2}\left(\pi T+\sqrt{\pi^2T^2-2c_2}\right)$). In the right plot of Fig.\ref{fig:diag}, we show the diagonal conductivities as a function of graviton mass for numerical results at $M/b=0.01$ and the analytical results at $M/b=0$. As the graviton mass is increased, the numerical results are equal to the analytical results. However, for large graviton mass, the numerical conductivities departure from the analytical conductivities, which is due to the appearance of spontaneous symmetry breaking solution discussed in the previous subsection. In summary, the behavior of diagonal conductivities under momentum relaxation is consistent with the phenomenon we found in anomalous Hall conductivity.

%%%%%%%%%%%%%%%%%%%%%%%%%
\section{Conclusion and discussion}\label{sec4}
%%%%%%%%%%%%%%%%%%%%%%%%%
In this work, we have studied the effects of momentum relaxation on the holographic Weyl semimetals, which is characterized by a quantum phase transition between the Weyl semimetal phase and a topological trivial phase. The momentum relaxation in field theory is induced by the non-zero graviton mass terms. By tuning the graviton mass, we have computed the anomalous Hall conductivity as a function of $M/b$, which further determines the critical value of the phase transition. We have found that the critical value decreases with the increasing of graviton mass and finally goes to zero above a critical graviton mass. This phenomenon is qualitatively similar to the results in Ref.\cite{Zhao:2021pih}, indicating a universal phenomenon in the momentum relaxed holographic Weyl semimetals, which is also consistent with the predictions from the weakly coupled field theory. We have also pointed out the existence of a spontaneous symmetry breaking solution at $M/b=0$, which interprets the deviation of anomalous Hall conductivity from $1$ near $M/b=0$ at large graviton mass.

\para
Notably, the physics described above has also evidenced from the behavior of diagonal conductivity, where the peak (minimal) in the transverse (longitudinal) conductivity goes to zero as the graviton mass is increased. We have also studied the diagonal conductivities as function of graviton mass at small $M/b$ and compared it with the analytical result at $M/b=0$.

\para
Our study of momentum relaxation in holographic Weyl semimetals reveals interesting and universal phenomenon of the system, and there are several directions that are worth the further study. First, it is natural to study the zero temperature ground-state of the holographic Weyl semimetals in the presence of non-zero graviton mass terms, which is important to reveal the underlying physics. Second, as the vector gauge field fluctuations studied in this paper decouple from the metric fluctuations, the d.c. conductivities are encoded in the equilibrium solutions. Thus, it will be interesting to study the axial gauge field fluctuations to see the effects of graviton mass at linear perturbation level.

%%%%%%%%%%%%%%%%%%%%%
\vspace{.8cm}
\subsection*{Acknowledgments}

We thank Yan Liu for useful discussion and guidance throughout the project. We thank Yan Liu and Hong-Da Lyu for reading the draft and providing helpful comments. We thank Hong-Da Lyu, Francisco Pena-Benitez and Hai-Qing Zhang for discussion. This work is supported by the National Natural Science Foundation of China Grant No.11875083.

\vspace{.3 cm}

%%%%%%%%%%%%%%%%%%%%%%%%%%%%%%%%
\appendix
\section{Equations of motion and asymptotic expansions}{\label{app:a}}
%%%%%%%%%%%%%%%%%%%%%%%%%%%%%%%%
Upon setting $2\kappa^2=L=1$, the background equations of motion for the ansatz (\ref{ansatz}) are
\bea
\frac{u''}{u}-\frac{f''}{f}+\frac{h'}{2h}\left(\frac{u'}{u}-\frac{f'}{f}\right)
     +\frac{m_g^2 \left(6F^2F_z c_3\sqrt{h}+2F F_z c_2\sqrt{f h}+2F^2c_2 h+F c_1\sqrt{f}h\right)}{u f h}&=&0  \,, \nn\\
\frac{u''}{2u}+\frac{f''}{f}+\frac{u'f'}{u f}-\frac{f'^2}{4f^2}-\frac{6}{u}-\frac{A_z'^2}{4h}+
    \frac{\phi^2}{2u}\left(m^2+\frac{\lambda}{2}\phi^2 -\frac{q^2A_z^2}{h}\right)
     +\frac{\phi'^2}{2} && \nn\\
     -\frac{m_g^2\left(F^2 c_2+F c_1\sqrt{f}\right)}{u f}&=&0  \,,  \nn\\
\frac{6}{u}-\frac{u'}{2u}\left(\frac{f'}{f}+\frac{h'}{2h}\right)-\frac{f'h'}{2fh}-\frac{f'^2}{4f^2}
     +\frac{A_z'^2}{4h}-\frac{\phi^2}{2u}\left(m^2+\frac{\lambda}{2}\phi^2+\frac{q^2A_z^2}{h}\right)+\frac{\phi'^2}{2} && \nn\\
     +\frac{m_g^2 \left(6F^2F_z c_3\sqrt{h}+4FF_z c_2\sqrt{f h}+2F^2 c_2h+2Fc_1\sqrt{f}h+F_z c_1f\sqrt{h} \right)}{2ufh}&=&0   \,, \nn\\
A_z''+\left(\frac{u'}{u}+\frac{f'}{f}-\frac{h'}{2 h}\right)A_z'-\frac{2q^2 \phi^2}{u}A_z&=&0  \,,  \nn\\
\phi''+\left(\frac{u'}{u}+\frac{f'}{f}+\frac{h'}{2h}\right)\phi'
     -\left(\frac{q^2 A_z^2}{uh}+\frac{m^2}{u}\right)\phi-\frac{\lambda \phi^3}{u}&=&0 \,, \nn
\eea
where the prime denotes the derivative of radial coordinate $r$. Note that, in contrast to the minimal model \cite{Landsteiner:2015pdh}, the above equations of motion contain new terms proportional to the graviton mass. Meanwhile, in the presence of non-zero graviton mass terms, the scaling symmetries involve the scaling of the reference metric. Particularly, there are three scaling symmetries
\\(I.)~~~$(x,y)\to\gamma(x,y),\,\, f\to \gamma^{-2}f,\,\, F \to \gamma^{-1}F \,; $ \\
(II.)~~$z\to\gamma z,\,\,  h\to \gamma^{-2}h, \,\,  (A_z, F_z)\to \gamma^{-1} (A_z, F_z)\,; $ \\
(III.)~$r\to\gamma r,\,\, (t,x,y,z)\to\gamma^{-1}(t,x,y,z), \,\, (u,f,h)\to\gamma^{2}(u,f,h), \,\,  (A_z, F, F_z)\to\gamma (A_z, F, F_z)\,; $ \\
%%%%%%%%%%%%%%%%%%%%%%%%%%%%%%%%
\subsection{Near horizon expansions}
%%%%%%%%%%%%%%%%%%%%%%%%%%%%%%%%
At the black hole horizon, we take the ansatz
\bea
u&=&4\pi T(r-r_h)+\cdots  \,, \nn\\
f&=&f_0+\Big[\frac{4F^3}{\sqrt{h_0}}m_g^2 c_3+(2F^2+\frac{2F^2\sqrt{f_0}}{\sqrt{h_0}})m_g^2c_2+
    (\frac{F f_0}{ 3\sqrt{h_0} }+\frac{5F\sqrt{f_0}}{3})m_g^2c_1 \nn\\
    &&+f_0(8-\frac{2m^2\phi_0^2}{3}-\frac{\lambda \phi_0^4}{3} )\Big] \frac{(r-r_h)}{4\pi T}+ \cdots  \,, \nn\\
h&=&h_0+\Big[\frac{4F^3\sqrt{h_0} }{f_0}m_g^2 c_3+\frac{4F^2\sqrt{h_0}}{\sqrt{f_0}}m_g^2c_2+
    (\frac{4F \sqrt{h_0} }{3}+\frac{2F h_0}{3 \sqrt{f_0} })m_g^2c_1  \nn\\
    &&+h_0(8-\frac{2m^2\phi_0^2}{3}-\frac{\lambda \phi_0^4}{3} )-2q^2A_{z0}^2\phi_0^2 \Big]\frac{(r-r_h)}{4\pi T}+\cdots   \,, \nn\\
A_z&=&A_{z0}+\frac{q^2A_{z0}\phi_0^2}{2\pi T}(r-r_h)+\cdots   \,, \nn\\
\phi&=&\phi_0+\frac{\phi_0 \big(q^2A_{z0}^2+h_0(m^2+\lambda \phi_0^2) \big)}{4h_0 \pi T}(r-r_h)+\cdots   \,, \nn
\eea
Note that, we have set $F_z=F$ in the above expansions. The independent parameters in the expansions are $T, r_h, f_0, h_0, A_{z0},\phi_0, c_i(i=1,2,3)$, which can be reduced by the above scaling symmetries and are mapped into dimensionless parameters ($\frac{M}{b}, \frac{T}{b}, \frac{m_gc_i}{b}$) in the field theory. Therefore, the numerical solutions can be obtained by integrating the background equations from the horizon to the AdS boundary by properly choosing the shooting parameters.

%%%%%%%%%%%%%%%%%%%%%%%%%%%%%%%%
\subsection{Asymptotic boundary expansions}
%%%%%%%%%%%%%%%%%%%%%%%%%%%%%%%%
Close to the conformal boundary, i.e. $r\to \infty$, we have
\bea
u&=&r^2+\frac{c_1Fm_g^2r}{3}-\frac{M^2}{3}+c_2F^2m_g^2+\frac{c_1Fm_g^2M^2+12c_3F^3 m_g^2}{6r}+
    \frac{(3\lambda+2) M^4 \ln r}{18r^2}+\frac{u_2}{r^2}\cdots  \,, \nn\\
f&=&r^2-\frac{M^2}{3}+\frac{8c_1F m_g^2M^2}{27r}
     +\Big[\frac{(3\lambda+2)M^4}{18}-\frac{2c_1^2 F^2m_g^4 M^2}{27}+\frac{c_2F^2m_g^2 M^2}{6} \Big]
     \frac{\ln r}{r^2}+\frac{f_2}{r^2}+\cdots   \,, \nn\\
h&=&r^2-\frac{M^2}{3}+\frac{8c_1F m_g^2M^2}{27r}
     +\Big[ \frac{(3\lambda+2)M^4}{18}+ \frac{q^2b^2 M^2}{2}
     -\frac{2c_1^2 F^2 m_g^4M^2}{27}+\frac{c_2F^2m_g^2M^2}{6} \Big]\frac{\ln r}{r^2}  \nn\\
     &&+\frac{h_2}{r^2}+\cdots  \,, \nn\\
A_z&=&b-bq^2M^2\frac{\ln r}{r^2}+\frac{\eta}{r^2}+\cdots  \,,  \nn\\
\phi&=&\frac{M}{r}-\frac{2c_1F m_g^2M}{3r^2}-
     \Big[ \frac{(3\lambda+2)M^3}{6}+\frac{M b^2 q^2}{2}-\frac{2c_1^2 F^2 m_g^4 M}{9}+\frac{c_2 F^2 m_g^2 M}{2}\Big]\frac{\ln r}{r^3}+\frac{O}{r^3}+\cdots  \,,  \nn
\eea
with $h_2=-2f_2+\frac{1}{8} b^2 M^2 q^2+\frac{7 M^4}{36}+\frac{\lambda M^4}{8}-M O-\frac{19}{54}c_1^2F^2 m_g^4M^2
+\frac{1}{8}c_2 F^2 m_g^2 M^2$. Note that, we have set $F_z=F$ to get this expansions. It is worth noting that the metric functions acquire nontrivial terms in the presence of non-zero graviton mass terms. In contrast to the minimal model of holographic Weyl semimetals \cite{Landsteiner:2015pdh}, the two conserved charges no longer exist under momentum relaxation. Therefore, we can not express $u_2, f_2$ and $h_2$ in the metric expansions in terms of $M, O, b$ and $\eta$. Furthermore, the above expansions are determined up to a radial shift $r\to r+a$.
%%%%%%%%%%%%%%%%%%%%%%%%%%%%%%%%
\section{Thermodynamics}\label{app:b}
%%%%%%%%%%%%%%%%%%%%%%%%%%%%%%%%
The Euclidean action reads
\bea
S_{\text{E} }=-\int d^5 x\sqrt{-g}\mathcal{L},
\eea
From the symmetry of the background ansatz (\ref{ansatz}), we can observe that the $tt$ component of $\chi_{ab}$ and $T_{ab}$ only contain terms proportional to the metric. Therefore, the $tt$ component of Einstein equation can be simplified as
\bea
R_{tt}=\frac{1}{2}g_{tt}\mathcal{L},
\eea
Therefore, the on shell action is a total derivative
\bea
S_{\text{E} }&=&-\int d^4 x \int_{r_h}^{r_\infty} dr \sqrt{-g}(2R^t_{\ t}) =\int d^4x\int_{r_h}^{r_\infty} dr [ u'f\sqrt{h} ]' \nn\\
    &=& \int d^4 x \bigg[ u'f\sqrt{h} \Big|_{r_\infty}-u'f\sqrt{h} \Big|_{r_h} \bigg],
\eea
Note that, the above expression is divergent near the AdS boundary. To compute the free energy, one need to add boundary counterterms, which takes the form
\bea
S_{ \text{ROS} }=S_{\text{E} }+S_{\text{GH} }+S_{\text{c.t.} } ,
\eea
where the Gibbons-Hawking term
\bea
S_{\text{GH} }=\frac{1}{\kappa^2} \int d^4 x \sqrt{-\gamma} K  \,, \nn
\eea
and the counterterm
\bea
S_{\text{c.t.} }&=&\frac{1}{2\kappa^2} \int d^4x \sqrt{-\gamma}\Bigg[-6-\Phi^2+
      \log(r)\bigg(\frac{1}{4}\mathcal{F}_{\mu\nu}\mathcal{F}^{\mu\nu}+
      \frac{1}{4}F_{\mu\nu}F^{\mu\nu}+|D_\mu\Phi|^2+\lambda_b \Phi^4 \bigg)  \nn\\
      &-&\frac{m_g^2c_1}{3} \mathcal{ \hat{U} }_1
      -\bigg(\frac{m_g^2c_2}{2}-\frac{m_g^4c_1^2}{72}\bigg) \mathcal{ \hat{U} }_2
      -\bigg(m_g^2c_3-\frac{m_g^4c_1c_2}{12}+\frac{m_g^6c_1^3}{432}
      -\frac{M^2m_g^2c_1}{12F^2}\bigg)\mathcal{ \hat{U}}_3 \Bigg] \,, \nn
\eea
Note that, $\lambda_b=\frac{1}{3}+\frac{\lambda}{2}-\frac{4 m_g^4c_1^2 F^2}{9M^2}+\frac{m_g^2c_2F^2}{M^2}$ and $\mathcal{ \hat{U}}_i (i=1,2,3)$ are graviton mass terms written in terms of induced metric $\gamma_{ij}$. The coefficients of each terms can be determined by demanding that the renormalized on-shell action is finite. Note that the counterterms of massive gravity herein are sufficient to cancel the divergence coming from the bulk graviton mass terms and see Refs.\cite{Chen:2019zlg} for more systematic treatment. Interestingly, in the presence of non-zero graviton mass terms, the counterterm of scalar field is slightly different.

\para
Having substituted the asymptotic expansions, we can obtain the renormalized on-shell action $S_{\text{ROS} }$. Consequently, the free energy reads
\bea
\frac{\Omega}{V}=-\frac{S_{\text{ROS}} }{V}&=&\frac{7 M^4}{36}-2 M O-T s-3u_2
                -\frac{23}{72}m_g^4 c_1^2 F^2 M^2 -\frac{1}{2} m_g^2 c_2 F^2 M^2 \nn\\
               &&+m_g^4 c_1 c_3 F^4+\frac{3}{4} m_g^4 c_2^2 F^4
               -\frac{1}{8} m_g^6 c_1^2 c_2 F^4+\frac{5 m_g^8 c_1^4 F^4}{1728}.
\eea
where $s=4\pi f_0\sqrt{h_0}$ is the entropy density.

\para
The expectation value of the boundary stress tensor reads
\bea
T_{\mu\nu}=2(K_{\mu\nu}-K \gamma_{\mu\nu})+\frac{2}{ \sqrt{-\gamma} }\frac{ \delta S_{\text{c.t.} } }{\delta \gamma^{\mu\nu}} \,,
\eea
Therefore, the energy density is
\bea
\epsilon=T^0_{\ 0}&=&\frac{7 M^4}{36}-2 M O-3u_2 -\frac{23}{72}m_g^4 c_1^2 F^2 M^2 -\frac{1}{2} m_g^2 c_2 F^2 M^2 \nn\\
               &&+m_g^4 c_1 c_3 F^4+\frac{3}{4} m_g^4 c_2^2 F^4
               -\frac{1}{8} m_g^6 c_1^2 c_2 F^4+\frac{5 m_g^8 c_1^4 F^4}{1728}.
\eea
Consequently, the thermodynamics relation $\frac{\Omega}{V}=\epsilon-Ts$ holds.

%%%%%%%%%%%%%%%%%%%%%%%%%%%%%

\end{document}